\documentclass[aps,pra,nofootinbib,preprint,amsmath,amssymb,floatfix]{revtex4-2}
\usepackage{color}
\usepackage{graphicx}
\usepackage{mathtools}

\begin{document}
\newcommand{\tc}{\textcolor}
\newcommand{\g}{blue}
\title{ Comment on the "Electric Power Generation from Earth's Rotation through its Own Magnetic Field"}

\author{Iver H. Brevik}      
\email{iver.h.brevik@ntnu.no}
\affiliation{Department of Energy and Process Engineering, Norwegian University of Science and Technology, N-7491 Trondheim, Norway}
\author{Moshe M. Chaichian}
\email{masud.chaichian@helsinki.fi}
\affiliation{Department of Physics, University of Helsinki, and Helsinki Institute of Physics,  P. O. Box 64, FI-00014 Helsinki, Finland}
\author{Mikhail I. Katsnelson}
\email{M.Katsnelson@science.ru.nl}
\affiliation{Institute for Molecules and Materials, Radboud University, Heijendaalseweg 135, 6525AJ, Nijmegen, Netherlands, and Wallenberg Initiative Materials Science for Sustainability (WISE), Uppsala University, Box 516, SE-75120, Uppsala, Sweden}

\date{\today}          

\begin{abstract}

The suggestion made by C. F. Chyba and K. P. Hand about electric power
generation from
Earth’s rotation through its own magnetic field is intriguing \cite{chyba16,chyba25}.
Due to the importance of the subject, we have re-analyzed the theoretical
arguments  and  derivations leading to  their conclusion, by  paying
special attention  to several  issues possibly neglected before. The model
they consider
is a magnetic cylindrical shell moving with  velocity $\mathbf{v}$ in the $y$
direction at a right angle to
the direction of the Earth’s magnetic field $\mathbf{B}_\infty$. First
we analyze the electromagnetic boundary conditions when the shell is
moving with a constant
velocity $\mathbf{v}$, as this point, although of importance,  has  not been taken
care of in \cite{chyba16,chyba25}.
 Indeed, this
procedure leads us to differences in the values of electromagnetic
fields when compared with the expressions
given in the cited references. Second  and as a result, we find that the
mechanical force created by the moving shell becomes different from the one  derived in \cite{chyba16,chyba25}.  Obviously,  the
expression for the  amount of electric power generation from Earth’s
rotation will also be different from the previously obtained one. The latter is
important for evaluating the amount of  produced  power, maximizing it
by choosing the parameters of the shell, and for the comparison with experimental
findings.

\end{abstract}
\maketitle

\bigskip
\section{Introduction}
\label{secintro}

\bigskip

The result obtained  by Chyba and Hand some years ago about generation of electric power
from Earth’s rotation through its own magnetic field has generated a lot of interest \cite{chyba16}, even
more so after the recent announcement about preliminary experimental results seemingly
supporting the theoretical predictions stated in \cite{chyba25}. As surveyed in
\cite{nature25}, physicists are divided
with respect to the  claim.

The geometrical model of the theory as well as the
model used in the experiments is that of a conducting  cylindrical long shell of inner radius $a$ and outer
radius $b$, where the material in the region $a\le\rho\le b$ is magnetic, with permeability $\mu$.
In the theoretical description of the system, a frame of reference is taken, where the shell is moving with a constant velocity $\mathbf{v} = v\hat{\mathbf{y}}$, $\mathbf{v}$ being the Earth rotation velocity at the place  of the shell  and considered to be linear, in a direction transverse to
the $x$ direction of the static magnetic field of the Earth, called $\mathbf{B}_\infty$. Thus, $\mathbf{B}_\infty = B_\infty \hat{\mathbf{x}}$. $\hat{\mathbf{x}}$ and $\hat{\mathbf{y}}$ are the unit vectors along the axes $x$ and $y$. The slight declination, at present 9.2 to 11.3 degrees, of the magnetic field relative to the geographic axis has been justifiably neglected in the estimates.

Some criticisms of this idea have appeared, cf., for instance, the paper of Jeener \cite{jeener20} and its  rebuttal in \cite{chyba20}. It is also  notable that in
an earlier experiment of Veltkamp and Wijngaarden, no energy production of this magnitude was found \cite{veltkamp18}. As expressed by Wijngaarden in \cite{nature25}: "I am still convinced that the theory of Chyba  {\it et al.} cannot be correct".

In view of the apparent uncertainty of the situation we have decided to
make a reconsideration
of the theory presented in \cite{chyba16} and \cite{chyba25}, although their theory is
detailed and very carefully
done. We shall  focus mainly on their first paper \cite{chyba16}. Evidently, there
is a missing piece
in their treatment, namely the electromagnetic boundary conditions on
the shell material surfaces
in motion. We discuss the implications of this omission for the main
conclusion of the
whole analysis, as our results are at variance with the once presented in the cited papers. We find
that our  expression
for mechanical  force  associated with the motion  is  different
compared with the one obtained in \cite{chyba16} and \cite{chyba25}. Obviously, the mechanical
work and produced power will also be different.

The experimental result  reported in \cite{chyba25} was very low, namely a continuous DC voltage of about $17~\mu$V, thus only of academic interest so far, but there might be a possibility for upscaling if the idea proves correct.

The experimental results reported in \cite{chyba25} will not be further considered here. To attempt a closer scrutiny of this would be complicated and lead us far away from the simple intention of the present note.

We start in the next section with the static case $\bf v=0$, and consider thereafter  the case where $ \bf v \neq 0$.

\section{The case $\bf v=0$}

The geometry of the setup is clearly sketched in \cite{chyba16} and \cite{chyba25}, and will not be reproduced here. It will however be convenient to reproduce the expressions for the components of the static magnetic field, called ${\bf B}_0$, both outside and inside  the conductive shell, in terms of cylindrical coordinates $\{ \rho, \phi, z \}$,
\begin{equation}
B_{0x}(\rho >b)= B_\infty +\beta_3(b/\rho)^2\cos 2\phi, \label{1}
\end{equation}
\begin{equation}
B_{0y}(\rho >b)= \beta_3(b/\rho)^2\sin 2\phi, \label{2}
\end{equation}
\begin{equation}
B_{0x}(a \leq \rho \leq b)=\beta_1-\beta_2(a/\rho)^2\cos 2\phi, \label{3}
\end{equation}
\begin{equation}
B_{0y}(a\leq \rho \leq b)= -\beta_2(a/\rho)^2 \sin 2\phi,
\end{equation}
\begin{equation}
B_{0x}(\rho \leq a)= 2\beta_1 (\mu_r+1)^{-1},
\end{equation}
\begin{equation}
B_{0y}(\rho \leq a) =0.
\end{equation}
Here, $y= \rho \sin \phi$ and the expressions for the coefficients are
\begin{equation}
\beta_1= 2B_\infty \mu_r(\mu_r+1)\zeta, \label{7}
\end{equation}
\begin{equation}
\beta_2= 2B_\infty \mu_r(\mu_r-1)\zeta,
\end{equation}
\begin{equation}
\beta_3= B_\infty [1-(a/b)^2](\mu_r^2-1)\zeta,
\end{equation}
\begin{equation}
\zeta= [(\mu_r+1)^2-(a/b)^2(\mu_r-1)^2]^{-1}.
\end{equation}
The magnetic vector potential is directed along the $z$ axis, ${\bf A_0}= A_0\hat{\bf z}$, where
\begin{equation}
A_0(\rho >b)= B_\infty y+\beta_3(b^2/\rho)\sin \phi, \label{A}
\end{equation}
\begin{equation}
A_0(a \leq \rho \leq b)=\beta_1y-\beta_2(a^2/\rho)\sin \phi,
\end{equation}
\begin{equation}
A_0(\rho <a)=2\beta_1(\mu_r+1)^{-1}\rho \sin \phi.
\end{equation}

\section{The case $ {\bf v \neq 0}$  }

 When the shell moves with constant velocity $\bf v$ in the $y$ direction with respect to  our rest frame\emph{}, the Maxwell equations within the shell will be modified. We shall call this shell-moving   frame of reference $K$, while the shell-at-rest reference frame  considered in the previous system is called $K_0$. Since the moving case  involves energy dissipation,  the conductivity $\sigma$ of the material will have to be accounted for.   In analogy to the well known Reynolds number in hydrodynamics, in  Refs. \cite{chyba16,chyba25}  was defined a magnetic Reynolds number, $R_m = v\xi/(\sigma \mu)^{-1}$, where $\xi$ is a typical length over which the current flow changes appreciably, and       $(\sigma \mu)^{-1}$ is the analog of the hydrodynamic kinematic viscosity. It is natural to associate $\xi$ with the outer radius $b$.

A central quantity in this kind of theory is the vector potential $\bf A$. It is pointing in the $z$ direction, ${\bf A}= A_z \hat{\bf z}$, in the frame $K$ as well as in $K_0$. The governing equation for $A_z$ in $K$ is (cf. Eq.~(25) in \cite{chyba16}),
\begin{equation}
\partial A_z/\partial t + v\partial A_z/\partial y = \eta \nabla^2A_z, \label{governingequation}
\end{equation}
 where $\eta = (\sigma \mu)^{-1}$ is conventionally called  magnetic diffusivity. The equation above is solved in two steps,
 \begin{equation}
 A_z= A_z(\rho, \phi) + A_t(\rho, \phi,t),
 \end{equation}
 where $A_z(\rho, \phi)$ relates to the steady state equation
 \begin{equation}
 v\partial A_z/\partial y = \eta \nabla^2 A_z, \label{16}
 \end{equation}
  and  $A_t(\rho, \phi,t)$ relates to the time-dependent equation
  \begin{equation}
  \partial A_t/\partial t + v\partial A_t/\partial y = \eta \nabla^2 A_t. \label{14}
  \end{equation}
 The solution found in  \cite{chyba16,chyba25} for the latter equation decays  exponentially in time,
 \begin{equation}
 A_t(\rho, \phi,t)= C_0\, m(\rho)n(\phi)\exp{(k\rho \sin \phi)}\cdot
\exp{[-\eta (k^2+\lambda^2)t]}.\label{timeequation}
 \end{equation}
 Here $C_0$ is a constant, $m(\rho)$ and $n(\phi)$ are linear combinations of Bessel functions and trigonometric functions respectively, and $k$ is defined as $k=v/2\eta$, thus a $K$-dependent quantity. The constant  $\lambda^2$ is related to a separation constant in Eq.~(\ref{14}). As emphasized  in  \cite{chyba16,chyba25}, the time-dependent part of the solution decays swiftly with the assumed low value of the Reynolds number $R_m$, and can be ignored when seeking for stationary solutions.

 The time-independent solution given, (Eq.~(49) in  \cite{chyba16}), is
 \begin{equation}
 A_z(a \leq \rho \leq b)= k\beta_1z^2+\beta_1y -\beta_2 ka^2K_1(k\rho)e^{ky}\sin \phi, \label{19a'}
 \end{equation}
 where $K_1$ is a Bessel function. From this the gradient of the scalar potential is derived as $\nabla V= -v\beta_1\hat{\bf z}$.  Correspondingly, the axial electric field in the interior of the shell, i.e. ${\bf E } = -\nabla V -\partial {\bf A}/\partial t$, becomes
 \begin{equation}
 {\bf E}(a \leq \rho \leq b)= v\beta_1\hat{\bf z}, \label{electricfield}
 \end{equation}
 when omitting the last term in (\ref{19a'}).
The electric field on the outside of the shell is obtained by setting $\mu_r=1$, whereby $\beta_1=B_\infty$, and so
\begin{equation}
{\bf E}( \rho > b)= v B_\infty \hat{\bf z}.\label{electricfieldoutside}
\end{equation}
There are two boundary conditions for dielectric boundaries moving with small velocities \cite{landau84}. When applied to $\rho =b$, they read
\begin{equation}
{\bf n}\times [{\bf E}(b+)-{\bf E}(b-)]= v_n [ {\bf B}(b+)-{\bf B}(b-)], \label{19}
\end{equation}
\begin{equation}
{\bf n}\times [{\bf H}(b+)-{\bf H}(b-)]= -v_n [ {\bf D}(b+)-{\bf D}(b-)], \label{19a}
\end{equation}
where $v_n$ is the velocity of the boundary along the normal to the boundary surface $\bf n$.

For a medium with zero conductivity, this boundary condition is straightforward. That the same condition holds also in the case of a general conductivity, when, as usual, the electric charges are distributed over the volume of the material without singular points or regions, follows from the Maxwell equations. First, Eq.~(\ref{19}), related to the rest-frame equation ${\bf \nabla \times E} = -\partial{\bf B}/\partial t$, does not contain $\bf J$ at all. Second, in Eq.~(\ref{19a}), related to ${\bf \nabla \times H }= {\bf J} + \partial {\bf D}/\partial t$, we can perform an integration over a coin-shaped volume whose main direction aligns with the surface. The volume contains the material surface.  Taking the limit when the thickness of the coin goes to zero, we see that the influence
from $\bf J$ has to vanish.  Only in exceptional cases containing singular currents, for instance, would $\bf J$ be able to contribute.

Inserting the expression (\ref{electricfield}) and (\ref{electricfieldoutside})  for $\bf E$  into the left-hand side  of (\ref{19}), we  get
\begin{equation}
[{\bf n \times \hat{\bf z}}]v(B_\infty -\beta_1)= v_n [ {\bf B}(b+)-{\bf B}(b-)].
\end{equation}
The fields $ {\bf B}(b+)$ and ${\bf B}(b-)$ are thus tangentially directed. Taking the $x$ components, we obtain
\begin{equation}
B_\infty -\beta_1= B_x(b+)-B_x(b-);
\end{equation}
the two  $v$ factors on each side drop out.

 We should now solve this equation with respect to the  component $B_x(b-)$ inside the surface. For that we also need  the expression for the exterior component $B_x(b+)$. This component  was not given explicitly in \cite{chyba16,chyba25},  but can be derived as follows, making use of the same method. We start from Eq.~(\ref{A}), modifying the last term, $A_0(\rho >b) = B_\infty y+h(z)$, where $h(z)$ is determined from Eq.~(\ref{16}) to be $h''(z)= 2kB_\infty,$ or $ h(z)= kB_\infty z^2.$ Going to the case $v \neq 0$, we get the extended form
\begin{equation}
A_z(\rho >b)= kB_\infty z^2+ B_\infty y + C\cdot K_1(k\rho)e^{ky},
\end{equation}
with $C$ a constant, and $K_1(k\rho)= 1/(k\rho)$ approximately for small values of $k\rho$.  The  value of $C$ is determined by requiring agreement with the expression (\ref{A}) for $v=0$ for small values of $ky$, implying $C=\beta_3kb^2$, leading to the external potential
\begin{equation}
A_z(\rho >b)= kB_\infty z^2+B_\infty y + \frac{\beta_3b^2}{\rho}\sin \phi (1+k\rho \sin\phi).
\end{equation}
Comparing this with the $v=0$ expression for $A_0$, we can write it in the form ($z^2$ term  omitted)
\begin{equation}
A_z(\rho >b)= A_0(\rho >b)+ \frac{1}{2}\beta_3 R_m b\sin\phi,
\end{equation}
where
\begin{equation}
R_m = \mu \sigma vb
\end{equation}
is the magnetic Reynolds number. In vacuum, $\sigma=0$, so we can just identify $A_z$ with $A_0$ on the outside.

Actually, this could have been seen in a more direct way, by exploiting the  gauge condition $ \nabla \cdot A=0$, which was found to apply also when $v \neq 0$ (cf. Eq.~(22) in \cite{chyba16} in the vacuum case $\eta \rightarrow \infty$).

Then we can make use of the expression (\ref{1}) directly in the frame $K$, implying that
\begin{equation}
B_{0x}(b+) = B_\infty + \beta_3 \cos 2\phi.
\end{equation}
Thus, the final result, in the same frame $K$, becomes
\begin{equation}
B_x(b-)= B_{0x}(b+)+\beta_1-B_\infty = \beta_1+\beta_3 \cos 2\phi. \label{final}
\end{equation}
This expression is independent of $v$.
It can be compared with the expression  presented in \cite{chyba16} (Eqs.~(59) and (60)),
\begin{equation}
B_x(b-)|_{CH}= B_{0x}(b+)- R_m \beta_2(a/b)^2\sin \phi \cos^2\phi, \label{CH}
\end{equation}
where, for low $v$,
\begin{equation}
 R_m =   2kb=     \mu \sigma v b.
 \end{equation}
Thus, our inclusion of the electromagnetic boundary conditions for moving boundaries leading to the expression (\ref{final}), disagrees with the expression (\ref{CH}) given in Ref. \cite{chyba16}.

\section{Produced power}

In this section we evaluate the produced power and also the
dissipation, although they have previously  been dealt with  in \cite{chyba16,chyba25}.
Due to our different values for the induced fields as compared with the
ones obtained in \cite{chyba16,chyba25}, the expression for the power and
dissipation will also be different. The  differences between the results
obtained  in \cite{chyba16,chyba25} and here all stem from the fact that we have taken
into account the requirement of boundary conditions. In the next
section we elaborate on the necessity and importance of boundary
conditions and illuminate it by mentioning a few known examples.

We consider the system  of reference in which the shell is moving.

Assuming that the medium has conductivity,  we start from Eq.~(\ref{electricfield}) which shows that in the frame K (in which the shell is moving) there is a constant electric field $\bf E = v\beta_1 \hat{\bf z}$ in the $z$ direction. There is thus an electrical force density component ${\rho\bf E}$ in that direction, $\rho$ being the charge density. This force yields  no work in the $y$ direction.

Secondly, we  consider  the magnetic part $\bf J\times B$ of the Lorentz force density, where  ${\bf J}=\sigma {\bf E}$ with $\sigma$ the conductivity. What expression is to be inserted for $\bf B$ here? The expression should be chosen such that the force reduces to zero in the static case.
So, when subtracting the latter fields, we obtain simply zero for the velocity-induced Lorentz force.

Of main interest is  the {\it induced} magnetic field   $ {\bf B}_{\rm induced}$. This field is calculated from a Galilean transformation between the two inertial systems as  $ {\bf B}_{\rm induced} = (1/c)^2{\bf v \times E}$ and lies in the $x$ direction.
 Thus, ${\bf J\times B}_{\rm induced}
= \sigma {\bf E}\times [\frac{1}{c^2}{\bf v\times E}]=
\sigma \beta E^2\hat{\bf y}/c$, where $\beta = v/c$. This force in principle does involve work in the $y$ direction. The corresponding power $P$ per unit volume and time is $P= \sigma \beta E^2 {\bf v\cdot \hat{y}}/c=\sigma \beta^2 E^2. $ Now, we make use of the order-of-magnitude relation $E \sim Bc$, which follows from Maxwell's equation ${\bf \nabla \times E}= -\partial {\bf B}/\partial t$ if we insert $|{\bf \nabla \times E}| \sim E/\lambda$, $|\partial {\bf B}/\partial t| \sim \omega B = (2\pi/\lambda)Bc$, with $\lambda$ the wavelength. One can alternatively express the power in terms of the magnetic field as $P= \sigma \beta^2B^2c^2$ (here $B$ meaning the induced field). As $\beta \ll 1$, the factor $\beta^2$ makes $P$ far too small.  By subtracting the static fields, we obtain an expression close to  zero. This is a main result in our investigation.


\section{Boundary conditions}

In our paper, an important point was the application of one of the  electromagnetic boundary conditions on dielectric surfaces in motion, using general formulas from Ref.~\cite{landau84}.  These are general conditions, that make no restriction on the magnitude of the conductivity $\sigma$. The physics of the boundary conditions is the following: In the inertial frame where the instantaneous velocity of the boundary is zero, the conventional conditions, namely the tangential component of the electric field $\bf E$ as well as   the tangential component of the magnetic  field $\bf H$, must be continuous. If they were not, this would lead to infinite surface charges or infinite surface currents at the boundary, and that is obviously not the case here.
These conditions, as we have  mentioned  above, are valid irrespective of whether the conductivity $\sigma$ has an  extremely low value (as is the case for a dielectric)  or is  finite for the case of a conductor shell, as far as $\sigma$ is a constant  independent of frequency.

To clarify still more the essence of the present Comment, its importance and differences with the original treatment in \cite{chyba16} and \cite{chyba25}, let us emphasize the following points:

In classical and also in quantum physics described by differential equations, without imposing boundary conditions one could add several arbitrary  (constants and constants multiplied by coordinates to some powers) terms to any solution of the differential equations and in that case obtain any value, including infinity,  for the physical quantities of the system under study, e.g. its total energy could become infinite.

A typical example in quantum mechanics is when we consider the solution of Schr\"{o}dinger's equation at the boundary of  a potential, e.g. a harmonic oscillator; one has to make sure that the wave function itself, as well as its first order derivative, are continuous at the boundary of the potential, which are  the matching conditions,  otherwise the equation would lose its meaning due to singularity and infinite value of the derivatives.

The solution of Maxwell's equations,  which is the main theoretical basis of  the present work  to start with, is the same and one should impose the boundary conditions, otherwise the solutions are not unique and the physical quantities can be arbitrary and different. This is the difference between the treatment of the present work and the original one performed in \cite{chyba16} and \cite{chyba25}

In the present work it is shown that the expressions for work performed
and for dissipation (although negligible)
   are different from the expressions obtained in the original treatment
performed in \cite{chyba16} and \cite{chyba25}.   Since the differences are in the form of
different  parameters of the shell, they are
important for experimental observations and for planning to maximize the
amount of electric power generation.

Another difference between the two treatments is that, while we consider the Maxwell equations and their solutions in a certain rest frame,  in \cite{chyba16} and \cite{chyba25}  they are considered in a moving system. However,  it is well known that the physical quantities are invariant and independent on the system of coordinates. Here, the work ${\bf F}\cdot{\bf r}$ is an invariant  scalar, since the system  in such a nonrelativistic motion of the Earth has symmetry under the Galilean transformations.

Therefore, the source of the discrepancy between the results of the present work and of the ones in \cite{chyba16} and \cite{chyba25}, comes only from the imposition of the boundary conditions.

\section{ Discussion and Summary}

For comparison,  as already mentioned, the effective value of the product ${\bf E\cdot J}$ obtained by Chyba and Hand is different from that we have found above. Writing an extra subscript CH to refer to their modified current density, we reproduce from Eq.~(81) in \cite{chyba16}:
\begin{equation}
{\bf E\cdot J}_{CH}= \sigma v^2\beta_1(\beta_1-B_x). \label{39}
\end{equation}
Here, the static contribution has been subtracted off. The power provided to the shell is found by integrating this expression over the volume.

The last equation (\ref{39}) for specific energy dissipation enables us to make an important observation. The magnetic field component $B_x$, referring to the inertial frame where the shell is moving, is given in the interior of the shell by Eqs.~(59) and (60) in Ref.~\cite{chyba16}, as the difference between the static field component $B_{0x}$  and a component $B_{1x}$. The latter is small, as is seen from the ratio $B_{1x}/B_{0x} \sim R_m = \mu \sigma vb$ which contains $v$. Thus, the contribution from   $B_{1x}$ to the power is negligible. What remains is the difference $\beta_1 -B_{0x}$, which is given by Eq.~(10a) in the same reference. This difference contains $\cos2\phi$\,: the expressions for $\beta_1$ and $B_{0x}$ are  given by eqs. (\ref{7}) and (\ref{3}), respectively.  Upon integration over $\phi$ from $0$ to $2\pi$ it vanishes, so that the total power given to the shell will be zero. That is barely an acceptable
global result. However, the local dissipation values, containing a cosine function, oscillate and become 
negative in a large region of $\phi$ around the shell, what is  physically not acceptable.

\section*{Acknowledgements}

We thank Markku Oksanen and Anca Tureanu for illuminating discussions and reading the manuscript,
and are grateful to Christopher F. Chyba, Kevin. P. Hand, and Thomas H. Chyba for valuable
comments on the first version of this paper.

\end{document}